\documentclass[conference,letterpaper]{IEEEtran}
\IEEEoverridecommandlockouts

\usepackage[utf8]{inputenc} 
\usepackage[T1]{fontenc}
\usepackage{amsmath,amssymb,amsfonts}
\usepackage{amsthm}
\usepackage{algorithmic}
\usepackage{graphicx}
\usepackage{textcomp}
\usepackage{xcolor}
\usepackage{comment}
\usepackage{cite}
\usepackage[inline]{enumitem}
\usepackage{bbm}
\usepackage[capitalize]{cleveref}
\usepackage{pgfplots}
\pgfplotsset{compat=newest}

\usepackage{tikz}
\usetikzlibrary{backgrounds,calc,shadings,shapes.arrows,shapes.symbols,shadows,fit,positioning,topaths,hobby}
\usepackage{balance}

\newcommand{\rtdesign}{\ensuremath{\kappa}}
\newcommand{\rtvar}{\kappa^\prime}

\newcommand{\romanenumitem}{\emph{(\roman*)}}

\newcommand{\E}[1]{\ensuremath{\mathbb{E}[#1]}}
\newcommand{\dist}{\ensuremath{\ell}}
\newcommand{\tdistrrgopt}[1][\dist]{\ensuremath{p_{\rtdesign, \rad}^\star(#1)}}
\newcommand{\tdistrrgoptplain}{\ensuremath{p_{\rtdesign, \rad}^\star}}

\newcommand{\tdistrrg}[1][\dist]{\ensuremath{p(#1)}}
\newcommand{\tdistrrgoptrad}[1][\dist]{\ensuremath{p_{\rad}^\star(#1)}}
\newcommand{\tdistrrgrad}[1][\dist]{\ensuremath{p_{\rad}(#1)}}

\renewcommand{\sp}{\ensuremath{\text{SP}}}
\newcommand{\nsp}{\ensuremath{\lnot \text{SP}}}
\newcommand{\retro}{\ensuremath{\text{RETRO}}}
\newcommand{\nretro}{\ensuremath{\lnot \text{RETRO}}}

\newcommand{\tfh}{\ensuremath{T_\text{FH}}}
\newcommand{\tfr}{\ensuremath{T_\text{FR}}}
\newcommand{\expnsp}{\ensuremath{\E{\tfh|\dist, \nsp}}}
\newcommand{\expnretro}{\ensuremath{\E{\tfr|\nretro}}}
\newcommand{\truncexpnsp}[1]{\ensuremath{\E{\tfh|\dist, \nsp, \tfh \leq #1}}}
\newcommand{\truncexp}[1]{\ensuremath{\E{\tfh|\dist, \tfh \leq #1}}}
\newcommand{\diam}{\ensuremath{d}}
\newcommand{\psp}{\ensuremath{\varphi}}

\newcommand{\distmin}{\ensuremath{\dist_1}}
\newcommand{\distmax}{\ensuremath{\dist_2}}

\newcommand{\V}{\mathcal{V}}
\newcommand{\Ed}{\mathcal{E}}

\newcommand{\partfac}{\ensuremath{K_{\rad}(\dist)}}

\newcommand{\fullfacrt}[1][\rtvar]{\ensuremath{E_{\rad, #1}(\dist)}}
\newcommand{\fullfacrtprime}[1][\rtvar]{\ensuremath{E_{\rad, #1}(\dist^{\prime})}}

\newcommand{\fullfacrtdistprime}[1][\rtvar]{\ensuremath{E_{\delta, #1}(\dist^\prime)}}

\newcommand{\cfac}{\ensuremath{c^\prime}}
\newcommand{\nnodes}{\ensuremath{N}}

\newcommand{\define}{\triangleq}

\newcommand{\errbound}[1][\rtvar]{\ensuremath{\varepsilon_\dist(\rtdesign, #1)}}
\newcommand{\errboundtmp}[1][\rtvar]{\ensuremath{\varepsilon^\prime_\dist(\rtdesign, #1)}}
\newcommand{\errboundtmpprime}[1][\rtvar]{\ensuremath{\varepsilon^\prime_{\dist^{\prime}}(\rtdesign, #1)}}

\newcommand{\iter}{g}

\newcommand{\commonsol}{s}
\newcommand{\rad}{\ensuremath{\delta}}
\newcommand{\pdfest}[1][\rtvar]{\ensuremath{W_{\delta, #1}}}
\newcommand{\prttails}{\delta^\prime}

\newcommand{\clientdata}[1][i]{\ensuremath{\mathcal{D}_{#1}}}

\newcommand{\modelsym}{\ensuremath{\mathbf{w}}}
\newcommand{\model}[1][\iter]{\mathbf{\modelsym}_{#1}}
\newcommand{\dimension}{d}

\newcommand{\lr}{\ensuremath{\eta}}
\newcommand{\neighbors}[1]{\mathcal{N}_{#1}}

\newcommand{\tv}[2]{\ensuremath{d_{\text{TV}}(#1\Vert #2)}}
\newcommand{\entropy}[1]{\ensuremath{\mathrm{H}\left(#1\right)}}
\newcommand{\binentropy}[1]{\ensuremath{\mathrm{h}_b\left(#1\right)}}

\newcommand{\avgitertime}{T_{\distmin, \distmax}}
\newcommand{\tveps}{\ensuremath{\rho}}

\newtheorem{theorem}{Theorem}
\newtheorem{corollary}{Corollary}
\newtheorem{lemma}{Lemma}

\newtheorem{proposition}{Proposition}
\newtheorem{definition}{Definition}

\newtheorem{assumption}{Assumption}

\newtheorem{remark}{Remark}

\begin{document}

\title{Source Anonymity for Private Random Walk Decentralized Learning}

\author{
\IEEEauthorblockN{Maximilian Egger, Svenja Lage, Rawad Bitar and Antonia Wachter-Zeh}
\IEEEauthorblockA{\IEEEauthorrefmark{1}%
                   Technical University of Munich, Germany, \{maximilian.egger, rawad.bitar, antonia.wachter-zeh\}@tum.de}
                   
\IEEEauthorblockA{\IEEEauthorrefmark{2}%
                   German Aerospace Center, Germany, \{svenja.lage@dlr.de\}} \vspace{-.5cm}
\thanks{This project is funded by DFG (German Research Foundation) projects under Grant Agreement Nos. BI 2492/1-1 and WA 3907/7-1.}
}

\maketitle

\begin{abstract}
This paper considers random walk-based decentralized learning, where at each iteration of the learning process, one user updates the model and sends it to a randomly chosen neighbor until a convergence criterion is met. Preserving data privacy is a central concern and open problem in decentralized learning. We propose a privacy-preserving algorithm based on public-key cryptography and anonymization. In this algorithm, the user updates the model and encrypts the result using a distant user's public key. The encrypted result is then transmitted through the network with the goal of reaching that specific user. The key idea is to hide the source's identity so that, when the destination user decrypts the result, it does not know who the source was. The challenge is to design a network-dependent probability distribution (at the source) over the potential destinations such that, from the receiver's perspective, all users have a similar likelihood of being the source. We introduce the problem and construct a scheme that provides anonymity with theoretical guarantees. We focus on random regular graphs to establish rigorous guarantees.
\end{abstract}

\section{Introduction}
Machine learning models have the potential to provide significant benefits in a wide range of areas, including intelligent healthcare~\cite{Gyorgy2024,Lian2022}, Internet of Things (IoT)~\cite{Messaoud2020} or Internet of Vehicles~\cite{Ali2021,Barbieri2022}. However, the success of the models relies heavily on access to large and comprehensive datasets. Distributed learning in its various forms, e.g., federated and decentralized learning, emerged as a new paradigm for accessing massive amounts of personalized and private data generated by participating clients. %

In federated learning~\cite{McMahan2016}, users maintain their data locally and only share locally updated models with the federator, who orchestrates the training process. Decentralized learning eliminates the need for a central authority. Instead, users take an active role in distributing model updates among themselves. The users can be modeled as vertices in a graph. Users who can communicate are connected with an edge. Two main types of algorithms are studied in the literature: \begin{enumerate*}[label=\emph{(\roman*)}]
    \item \emph{gossip algorithms}, e.g., \cite{Boyd_2005, Shah_2009,Lu2010GossipAF,Boyd_2006,Nedic2009DistributedSM,koloskova2019decentralized,duchi2011dual}, in which at every iteration, all users update the model locally and share their update with all their neighbors; and \item \emph{random walk-based algorithms}, e.g., \cite{Johansson2009ARI,ayache2023walk,duchi2012ergodic,ayache2021,egger2024self}, in which at every iteration, one designated user updates the model locally and shares the update with one of its neighbors chosen at random.
\end{enumerate*} In both cases, the algorithm proceeds until certain convergence criteria are met.
We focus on random walk-based algorithms due to their low communication cost incurred per iteration. 
The name random walk-based algorithm stems from the machine learning model being passed sequentially among neighboring nodes, thus drawing a random walk (RW) on the graph. 

Despite users' data being kept locally, privacy is not immediately preserved. For instance, users can infer updates to the model by comparing its state to the point where the RW was last observed. The current model might exhibit a bias towards the data of the user who recently updated it. By accumulating such observations, a user may potentially glean information about the other users' data, cf. \cite{Yin2021,Hu2022}. 

The main approach to conceal individual data updates is through the application of differential privacy \cite{Dwork2014}, as done in~\cite{ayache2021}, by injecting carefully designed random noise into the model updates. However, this comes at the cost of a trade-off between privacy and model precision~\cite{Mengqian2021}. The noise needed to ensure privacy grows with the number of updates that will be observed \cite{Kairouz_2015}. Therefore, without proper care, a large amount of noise may be needed, significantly reducing the algorithm's utility. While homomorphic encryption, as introduced in \cite{Rivest1978}, presents a promising alternative as it allows for computation on encrypted data, the computational overhead of such algorithms often renders them impractical for handling large datasets or complex functions (see, for example \cite{Lee2022}).

This paper introduces a novel privacy-preserving model that does not require differential privacy mechanisms and whose practicality surpasses homomorphic solutions. 
The core idea is to use public-key encryption and source anonymity as follows. The user updating the model encrypts it using the public key of the destination user. The model is transmitted through the graph in a way that, when reaching the destination, the identity of the transmitting user will be concealed. This model achieves two goals simultaneously: \begin{enumerate*}[label=\romanenumitem]
    \item \emph{high utility}, the destination can decrypt the model and use it plainly; and \item \emph{privacy}, by concealing the identity of the transmitter, the eavesdropping users would not be able to map the information they inferred to other users' data.
\end{enumerate*}
Note that our privacy-preserving mechanism is compatible with differential privacy. Noise can be inserted into the model before encryption. The noise level needed here may be lower since concealing the identity of the user hinders the composition of multiple observations.

The main challenge in this model is to carefully design a choice of the destination node among all nodes to ensure that the identity of the transmitting user is concealed. We term this property ``source anonymity'', inspired by similar studies in other contexts, such as wireless sensor networks, e.g., \cite{alomair2010statistical,yang2013towards}. To the best of our knowledge, this work is the first to tackle anonymity for random walk-based learning. We investigate random regular graphs in which a rigorous analysis of our method can be carried out. We study the privacy leakage, i.e., how well the source anonymity can be guaranteed under probabilistic guarantees. %

\section{Model and Anonymity Notion}
Consider a collaborative setup consisting of $\nnodes$ users, also referred to as nodes. Each user $i$ with $i\in [\nnodes]\define \{1,\dots,\nnodes\}$, possesses a local dataset $\clientdata$ and is capable of communicating with a subset of the other users $\neighbors{i} \subseteq [\nnodes]$, called its neighbors. We represent the users as vertices in a graph $\mathcal{G} = (\V,\Ed)$, where the set of vertices $\V = [\nnodes]$ corresponds to the users, and the edges $\Ed$ represent the communication links between them. The degree of a node $i \in \V$ is defined as the number of neighbors it has, i.e., $\deg(i) = |\{j\in\V: (i, j) \in \Ed\}| = |\neighbors{i}|$.

We sketch a potential learning algorithm that serves as motivation for our work, noting that our methods can be applied to any RW-based decentralized system that processes sensitive individual data. Starting with a random model $\model[0]\in \mathbb{R}^\dimension$ and a designated node $i_0\in \V$, the following procedure takes place at every iteration $\iter\geq 0$. Node $i_\iter$ updates the model using (stochastic) gradient descent based on a predefined loss function $F(\model[], \mathcal{D}_{i_\iter})$ on its local data $\mathcal{D}_{i_\iter}$, i.e., the model update reads $\model[\iter+1] = \model[\iter] - \lr \nabla F(\model[\iter], \mathcal{D}_{i_\iter})$. The updated model is then sent to a random neighbor $i_{\iter+1} \in \neighbors{i_\iter}$. The process repeats until certain convergence criteria are met. 

The described learning approach implicitly leaks information about the nodes' private data through the shared model updates. The privacy problem becomes most pronounced when eavesdropping nodes can obtain information about the model update of the targeted neighbors. For instance, consider a situation where, at iteration $\iter$, a node $i_{\iter}$ sends the model update to a neighbor $i_{\iter+1}\in \neighbors{i_\iter} $, who updates the model and then chooses $i_{\iter+2} = i_\iter \in \neighbors{i_{\iter+1}}$ as the next destination. In this case, node $i_\iter$ can directly obtain the local model update (gradient) $\nabla F(\model[\iter+1],  \clientdata[\protect{i_{\iter+1}}])$ of node $i_{\iter+1}$, and from this infer information about node $i_{\iter+1}$'s local data $\clientdata[i_{\iter+1}]$.

We propose a modification of this learning algorithm that does not allow model updates conducted by direct members and further hides the identity of the updating node. In particular, node $i_{\iter+1}$ should not know the identity of the predecessor $i_\iter$, thereby providing a new notion of privacy through source anonymity. We focus on a single iteration in the following and hence drop the iteration index $\iter$. 

We assume that each node generates a cryptographic key pair and publishes its public key. Upon updating the model, the current node $i\in\V$ selects a destination node $j \in \V\setminus\{i\}$ according to a predefined distribution $p_{D_i}$. This distribution may be both node-dependent and time-varying. The current node then encrypts the model using the public key of the destination node $j$. The RW continues independently moving on the graph, but no update is performed until the destination node is reached. Once the RW reaches node $j$, the model can be decrypted using node $j$'s private key. After updating the model, the learning process proceeds. The time it takes for the RW to transition from node $i$ to node $j$ is known as the first hitting time, denoted by $\tfh(i,j)$. We define the first hitting time distribution as $p_{i \rightarrow j} (t)$, which represents the probability that the RW reaches node $j$ for the first time at time $t$, starting from node $i$. The return time $\tfr(i)$ is the time it takes for an RW to return to a node $i$ after leaving the node. 

By encrypting the model for a designated destination node and choosing appropriate distributions $p_{D_i}$, we ensure that the destination node cannot determine the identity of the source node, thereby preserving the source node's anonymity. Our formal privacy notion is defined as follows.

\begin{definition}\label{PrivacyNotion}
    Let $H$ denote the entropy function. %
    We say that an RW-learning model, as described above, ensures $\alpha$-privacy if, for intervals $I_{i,j}\define[E_{i,j}-\rad,E_{i,j}+\rad]$ and fixed $\rad>0$, %
    \begin{align}\label{Gl1}
    \min_{j\in \V} H\Bigg( \left[ \frac{\int_{I_{i,j}} p_{D_i}(j) p_{i \rightarrow j} (t) dt}{\sum_{i^{\prime}\in\V\setminus\{j\}} \int_{I_{i^{\prime},j}} p_{D_{i^{\prime}}}(j) p_{i^{\prime} \rightarrow j} (t) dt} \right]_{i\in\V\setminus\{j\}} \Bigg) \geq \alpha,\\[-1cm] \nonumber
    \end{align}
   where $\alpha>0$ and $E_{i,j}\geq\delta$ for all $i,j\in\V$.
\end{definition}

Imagine that node $j\in\V$ receives the RW and attempts to estimate the identity of the source node. To facilitate this, node $j$ estimates the time interval for an RW to move from a source node to itself, denoted as $E_{i,j}$. This estimation can be accomplished through various means, including computing average values or incorporating supplementary side information. For every node $i\in\V\setminus\{j\}$, and an estimated path length around $E_{i,j}$, node $j$ calculates the probability that node $i$ was the sender. If the entropy over all possible source nodes is high, node $j$ will struggle to distinguish the true source node from other nodes. In the optimal scenario, the possible source nodes appear uniformly distributed to the destination node, making it impossible to identify the true source. Finally, note that this condition must hold for every possible destination node $j\in \V$.

Concealing the source of an update incurs a cost in terms of increased runtime. Unlike the classical approach, where the model is updated at every time step, our method updates the model only when the RW reaches the designated destination node. On average, the first hitting time is given by 
\begin{align*}
    \E\tfh=\sum_{i,j\in \V, i\neq j} \mathbb{E}[\tfh(i,j)]P(i)p_{D_i}(j), 
\end{align*}
where $P(i)$ represents the likelihood of the RW being in node $i\in\V$. In practice, choosing suitable distributions $p_{D_i}$ requires balancing two competing goals: achieving a high level of anonymity while also minimizing the runtime overhead.

\section{Optimal Source Anonymity} \label{sec:preliminary_solution_rrg}
To elucidate the implications of the privacy notation outlined in the previous section, we examine the specific instance of Random Regular Graphs (RRG). The uniform structure of RRGs enables a thorough and precise analysis, leveraging analytical expressions for the first hitting time and return time of the RWs. This characteristic makes RRGs a more accessible and illustrative choice for demonstrating our model, compared to other, more complex random graphs. 
We show in this section how to select the distributions $p_{D_i}$ such that, in the absence of additional side information, our model yields an optimal privacy guarantee; i.e., that the potential source nodes appear uniformly distributed from the perspective of the destination node. We therefore give a concise introduction to RRGs and their associated first hitting time distribution. We then address the topic of source node anonymity within this context. 

\subsection{On the First Hitting Time of RRGs}
RRGs are characterized by a degree distribution, that, for all nodes, is a degenerate probability density function, such that $P(k) = \mathbbm{1}_{\{c\}}(k)$, where $c$ is the degree of the RRG, and $\mathbbm{1}_{\{c\}}(k)=1$ iff $c=k$ and $0$ else. Hence, every node within the graph exhibits a uniform degree. We focus on RRGs in which the degree parameter $c\geq 3$, for which the graph consists of a single, connected component if $N$ is large enough \cite{Bollobás2001}.

For RRGs, the authors of \cite{Tishby2022} presented approximate expressions for the first hitting time distribution. Notably, they considered two distinct cases for the RW between two nodes.  In the shortest path ($\sp$) scenario, the RW follows the direct path between node $i$ and node $j$. This includes scenarios in which the RW may backtrack some of its steps or even recede. Formally, a path belongs to this case if the subnetwork consisting of the nodes and edges along the trajectory forms a tree network, and the distance between node $i$ and node $j$ in this subnetwork is the same as in the entire network. All other paths belong to the opposite case, denoted as $\nsp$.

We are interested in the first hitting time between nodes $i$ and $j$, which, in an RRG, only depends on their distance $\dist$ (i.e., the length of the shortest path). According to \cite{Tishby2022}, we have
$
    P(\tfh=t\vert \dist) = P(\tfh=t\vert \dist,\sp)P(\sp\vert \dist) + P(\tfh=t\vert \dist, \nsp)P(\nsp\vert \dist), 
$
where the hitting time distributions conditioned on the two distinct cases $\sp$ and $\nsp$ are given by $ P(\tfh=t\vert \dist,\sp)  = 
\frac{\dist}{t} \binom{t}{\frac{t+\dist}{2}} (1-c^{-1})^{\frac{t+\dist}{2}} c^{\frac{\dist-t}{2}}$ for $(t-\dist)$ even and
\begin{align}
        P(\tfh=t\vert \dist, \nsp)& =
          \left(e^{\frac{\cfac}{\nnodes}}-1\right) e^{-\cfac\frac{t-\dist}{\nnodes}} \label{Gl4},
\end{align}
for $t>\dist$, where $\cfac$ is defined as $\cfac \define (c-2)/(c-1)$. The probabilities for each case are given by
\begin{align*}
           P(\sp\vert \dist) = \left(\frac{1}{c-1}\right)^\dist +\frac{1}{\nnodes} & \text{ and }
    P(\nsp\vert \dist) = 1- P(\sp\vert \dist). 
\end{align*}

Additionally, we can express the expected value of the first hitting time distribution, conditioned on the distance between two nodes, as $\mathbb{E}[\tfh\vert\dist] =\E{\tfh\vert \dist,\sp}P(\sp\vert \dist)+ \E{\tfh\vert \dist,\nsp}P(\nsp\vert \dist)$
with $\E{\tfh\vert \dist,\sp}  = \frac{c}{c-2}\dist$ and
\begin{align}\label{Gl3}
    \E{\tfh\vert \dist,\nsp} & = \dist +\frac{1}{1-e^{-\frac{\cfac}{\nnodes}}}.
\end{align}

In the non-shortest path scenario, we applied the more accurate result \cite[Eq. 54]{Tishby2022} for $P(\tfh>t\vert \dist, \nsp)$ to calculate the expectation, yielding a result that differs marginally from the original expression in \cite[Eq. 70]{Tishby2022}.

\subsection{Source Anonymity in Light of RRGs}
We demonstrate how to select the distributions $p_{D_i}$ to achieve optimal $\alpha$-privacy, where optimal refers to a value of $\alpha$ equal to the entropy of a uniform distribution. For simplicity, we assume that all distributions $p_{D_i}$ are time-invariant. Furthermore, due to the regular structure of the graph, we posit that for an arbitrary node $i$, the probability distribution $p_{D_i}(j)$ depends solely on the distance between nodes $i$ and $j$. Let $A(\dist)$ be the number of nodes situated at distance $\dist\geq 1$, which we assume is constant for every node~$i$. We can then equivalently express the distribution $p_{D_i}$ as $p_{D_i}(j) = \frac{\tdistrrg }{A(\dist)}$, where $p$ denotes a probability distribution over node distances. To achieve high-probability control in arbitrary cases, we constrain the support of $p$ by excluding direct neighbors. Denote the resulting support of $p$ as $[\distmin,\distmax]$, where $\distmax$ is bounded from above by the diameter of the graph.

Initially, we consider a scenario in which the node lacks any supplementary side information beyond the graph's structure and the distributions $(p_{D_i})_{i\in\V}$. Under these conditions, a natural choice for $E_{i,j}$ in (\ref{Gl1}) is the expected first hitting time  $E_{i,j}=\mathbb{E}[\tfh(i,j)]$ from node $i$ to node $j$. Notably, in an RRG, the expected first hitting time $E_{i,j}$ depends solely on the nodes' distance $\dist$. The integral in the numerator of (\ref{Gl1}) simplifies to 
\begin{align}\label{Gl2}
    \int_{\mathbb{E}[\tfh(i,j)]-\rad}^{\mathbb{E}[\tfh(i,j)]+\rad} \hspace{-0.9cm} p_{D_i}(j) p_{i \rightarrow j} (t) dt
    &\!=\!\frac{\tdistrrg }{A(\dist)} \hspace{-0.1cm} \sum_{t=\mathbb{E}[\tfh\vert \dist]-\rad}^{\mathbb{E}[\tfh\vert \dist]+\rad} \hspace{-0.5cm} P\left(\tfh = t\vert \dist\right). 
\end{align}

When computing the sum in (\ref{Gl2}), we focus on the hitting time probability within an interval centered around the expected time.  However, as a consequence of (\ref{Gl3}), it follows that
$
    \mathbb{E}[\tfh\vert\dist] \geq \E{\tfh\vert \dist,\nsp}P(\nsp\vert \dist)
    \sim \left(l+\frac{N}{\cfac}\right)P(\nsp\vert \dist)
$
and hence, $\mathbb{E}[\tfh\vert\dist]\gg l$. For values of $t$ in this regime, we have 
$
     P(\tfh=t\vert \dist,\sp) \in O(t^{-\frac{3}{2}}e^{-\gamma t})
$
with 
$
    \gamma = \log(c)-\frac{1}{2}\log(c-1)-\log(2)>0,
$
which, for all $c$ becomes asymptotically negligible compared to the contribution from the non-shortest path. This observation justifies the following simplification.

\begin{assumption} \label{ass:fht}
    The first hitting time distribution within the interval $I_\dist\define[\mathbb{E}[\tfh\vert\dist]-\rad,\mathbb{E}[\tfh\vert\dist]+\rad]$ is dominated by the non-shortest path scenario for all $ \dist \in [\distmin, \distmax]$, i.e., we assume that $\Pr(\tfh = t \vert \dist) \define \Pr(\tfh = t \vert \dist, \nsp) \Pr(\nsp \vert \dist)$ for every $t\in I_\dist, \dist \in [\distmin, \distmax]$.
\end{assumption}
Under this assumption, the integral presented in (\ref{Gl2}) admits a closed-form solution. To simplify the notation, let
\begin{align*}
    \partfac \define \frac{P(\nsp\vert \dist)}{A(\dist)}\left(e^{\frac{\cfac}{\nnodes}}-1\right) e^{\cfac \frac{\rad+\dist}{\nnodes}} \frac{e^{-\cfac\frac{2\rad+1}{\nnodes}}-1}{e^{-\frac{\cfac}{\nnodes}}-1}.
\end{align*}

\begin{lemma} \label{lem:unnormalized_likelihood_sp}
Let $i,j\in \V$ be two nodes within distance $\dist\in[\distmin,\distmax]$ and let $\delta>0$ be fixed. Under \cref{ass:fht}, the integral in (\ref{Gl2}) is given by  
\begin{align*}
    &\int_{\mathbb{E}[\tfh\vert\dist]-\rad}^{\mathbb{E}[\tfh\vert\dist]+\rad} p_{D_i}(j) p_{i \rightarrow j} (t) dt = \tdistrrg E_{\delta}(\dist)
\end{align*}
where $E_{\delta}(\dist) = \partfac e^{-\cfac\frac{\mathbb{E}[\tfh\vert\dist]}{\nnodes}}$.
\end{lemma}

We find that the privacy notion now only depends on the distance $\dist$ between two nodes. Consequently, in the context of RRGs, our privacy notion requires that the distribution 
\begin{align*}
    W_{\delta}(\dist) \define \frac{p(\dist) E_{\delta}(\dist)}{\sum_{\dist^{\prime}} p(\dist^{\prime})E_{\delta}(\dist^{\prime})}
\end{align*}
on $\dist\in[\distmin,\distmax]$ remains sufficiently close to a uniform distribution over all possible values of $\dist$. Notably, this security notion is particularly satisfied when $p(\dist) E_{\delta}(\dist)$ remains constant across all values of $\dist \in [\distmin, \distmax]$. The following choice of $p(\dist)$ achieves optimal privacy in the sense of (\ref{Gl1}) for RRGs.

\vphantom{test}
\vspace{-1.55cm}
\begin{lemma}\label{lemma:rrgoptprelim_sp}
    \vspace{1cm}Let $\delta>0$. For every $\dist\in[\distmin,\distmax]$, choose $p(\dist)$ as
    \begin{align*}
        \tdistrrgoptrad %
        \define\frac{1}{E_{\delta}(\dist)\sum_{\dist^\prime = \distmin}^{\distmax} \frac{1}{E_{\delta}(\dist^{\prime})}}
    \end{align*}
 and $\tdistrrgoptrad= 0$  elsewhere. Then $W_{\delta}$ is uniform on the support $[\distmin, \distmax]$ and consequently $H(W_{\delta}) = \log(\distmax-\distmin+1)$. 
\end{lemma}

This choice of $p(\dist)$ guarantees the destination node cannot identify the source node with better accuracy than by simply choosing uniformly at random from all possible nodes. This demonstrates how to achieve optimal $\alpha$-privacy with the maximal value of $\alpha$, if no further side information is available.

\begin{remark}
    For RRGs, we analyze the distributions over distances. Equivalently, the entropy can be formulated expanding each distance $\dist$ with all source nodes in distance $\dist$.
\end{remark}

\section{Source Anonymity under Side Information}
In practice, the situation is more complex than initially described. A node can gather side information about the first hitting time by recalling the most recent visit of the RW. This additional information can, in turn, affect the node's ability to accurately infer the identity of the source node. This consideration becomes particularly crucial when the return time between two visits of the RW is short, as it effectively eliminates certain nodes as potential sources. If the model was re-encrypted between two returns to node $j$, and node $j$ is the designated destination, the return time represents an upper bound on the first hitting time. If the distributions $p_{D_i}$ were chosen as before, the destination node could indeed make a more informed guess about the source node, surpassing the uniform case. To alleviate this problem, users can choose $p_{D_i}$ such that even with this additional side information, the deviation from the uniform distribution remains bounded with high probability.
To optimize the model for such cases, the user selects a design parameter $\rtdesign$ as an upper bound for the first hitting time. Following this, $E_{i,j}$ in Definition \ref{PrivacyNotion} can be chosen as $E_{i,j} = \mathbb{E}[\tfh\vert \dist, \tfh \leq \rtdesign]$. We will first demonstrate how to choose $p_{D_i}$ in this setting and then analyze the probabilistic guarantees when a different value $\rtvar$ is observed during a random walk, where $\rtvar$ refers to the return of the random walk used as side information by the destination node as described above. In line with Assumption \ref{ass:fht}, we assume the following.

\begin{assumption}\label{ass:eth}
    $\truncexp{\rtvar}$ is dominated by the non-shortest path scenario, that is $\truncexp{\rtvar}\approx\truncexpnsp{\rtvar} \Pr(\nsp \vert \dist, \tfh \leq \rtvar)$.
\end{assumption}
The expected value obtained under additional side information can be calculated as follows. 
\begin{proposition} \label{prop:truncexpfht}
For every $\rtvar>\dist$, we have 
\begin{align*}
    \truncexpnsp{\rtvar} = \frac{\dist}{1 - e^{-\cfac/\nnodes (\rtvar-\dist)}} + \frac{1}{1-e^{-\cfac/\nnodes}}.
\end{align*}
\end{proposition}

The following generalization of \cref{lem:unnormalized_likelihood_sp} gives a closed-form expression for the corresponding modification of \eqref{Gl2} with additional side information. 

\begin{lemma} \label{lem:unnormalized_likelihood}
Let $i,j\in \V$ be two nodes within distance $\dist\in[\distmin,\distmax]$. Let $\delta>0$ be fixed and assume that it is known that the first hitting time is restricted by $\rtvar\in(\dist,\infty)$. Under \cref{ass:fht}, we have  
\begin{align*}
    &\int_{\truncexp{\rtvar}-\rad}^{\truncexp{\rtvar}+\rad} p_{D_i}(j) p_{i \rightarrow j} (t) dt = \tdistrrg \fullfacrt[\rtvar],
\end{align*}
where $\fullfacrt[\rtvar] = \partfac e^{-\cfac\frac{\truncexp{\rtvar}}{\nnodes}}$.
\end{lemma}

We refine the solution given in \cref{lemma:rrgoptprelim_sp} for the optimal distribution of $\tdistrrgoptrad$ to the case where certain side information $\rtvar = \rtdesign$ is assumed. Afterwards, we provide guarantees on privacy when the actual side information $\rtvar$ differs from the design parameter $\rtdesign$. Given $\rtdesign$, we choose the destination node distributions as follows.
\begin{lemma} \label{lemma:rrgopt}
Let $\rtdesign, \rad > 0$. For every $\dist \! \in \! [\distmin,\distmax]$, choose $p(\dist)$ as
\begin{align*}
        \tdistrrgopt =\frac{\commonsol}{\fullfacrt[\rtdesign]}=\frac{1}{\fullfacrt[\rtdesign] \sum_{\dist^\prime = \distmin}^{\distmax} \frac{1}{\fullfacrtdistprime[\rtdesign]}}, %
\end{align*}
and $\tdistrrgopt[\dist] = 0$ elsewhere. With $\pdfest[\rtvar]$ defined as  
\begin{align}
    \pdfest(\dist) \define \frac{\tdistrrgopt \fullfacrt[\rtvar]}{\sum_{\dist^{\prime}} \tdistrrgopt[\dist^{\prime}] \fullfacrtprime[\rtvar]}, \label{eq:designdist}
\end{align}
we have that $\pdfest[\rtdesign]$, i.e., $\pdfest(\dist)$ for $\rtvar=\rtdesign$, is uniform on the support $[\distmin, \distmax]$ and consequently $\entropy{\pdfest[\rtdesign]} = \log(\distmax-\distmin+1)$. %
\end{lemma}

Let $\tdistrrgopt$ be the optimal distribution for the case $\rtvar=\rtdesign$. To analyze the privacy for $\rtvar \neq \rtdesign$, we study the distribution $\pdfest(\dist)$,
which inherits the support $[\distmin,\distmax]$ from $\tdistrrgoptplain$. Our objective is to show that, with high probability, the deviation between $\pdfest$ and the uniform distribution of $\pdfest[\rtdesign]$ is bounded, facilitating a bound for the $\alpha$-privacy guarantee. Using \cref{prop:truncexpfht}, we can bound the total variation and the entropy as follows.

\begin{theorem} \label{thm:tvboundest}
Let $\rtdesign, \rtvar, \delta > 0$. For $\pdfest$ as in \eqref{eq:designdist}, we have
\begin{align*}
    \tv{\pdfest}{\pdfest[\rtdesign]} \leq \frac{1}{\distmax-\distmin} \sum_{\dist=\distmin}^{\distmax} (e^{\frac{\cfac}{\nnodes} \psp \errbound} - 1),
\end{align*}
where $\errbound = \dist \left\vert \frac{1}{1 - e^{-\cfac/\nnodes (\rtvar-\dist)}}-\frac{1}{1 - e^{-\cfac/\nnodes (\rtdesign-\dist)}}\right\vert$, and $\psp \define \Pr(\nsp\vert \distmax, \tfh \leq \rtvar)$.
\end{theorem}

\begin{theorem} \label{thm:estentropybound}
Let $\tveps \define \tv{\pdfest}{\pdfest[\rtdesign]}$. For $\tdistrrgopt$ as in \cref{eq:designdist}, the entropy of $\pdfest[\rtvar]$ is lower bounded by
\begin{align*}
    \entropy{\pdfest[\rtvar]} \geq (1-\tveps) \log(\distmax-\distmin+1) + \tveps \log(\tveps) - \tveps.
\end{align*}
\end{theorem}
\begin{figure}[t]
    \vspace{.2cm}
    \centering
    \resizebox{.99\linewidth}{!}{\input{entropies_vs_trt_0.3_4_300_5}}
    \vspace{-0.25cm}
    \caption{RW-model for $\nnodes=300$ nodes and degree $c=3$ optimized for $\rtdesign=634$ and $\rad=5$. We compare the entropies over varying side information $\rtvar$, with and without \cref{ass:fht}, compared with no countermeasure and the theoretical guarantee from \cref{thm:optimalsol_entropybound} using $\prttails=0.3$. Uniform distribution supported on $[\distmin, \distmax] = [2,6]$ has entropy $\log(\distmax-\distmin+1)\approx 1.61$.}
    \vspace{-.3cm}
    \label{fig:leakage_over_sideinformation}
\end{figure}

The distribution of $\rtvar$ is captured by the first return time of the RW. The deliberate choice of the destination node does not affect the stochasticity of the RW, and hence the return time is independent of our method. Let $\mathcal{K} \subset \mathbb{N}$ be such that $\quad \Pr(\rtvar \in \mathcal{K}) \geq 1-\prttails$ Our goal is that for a given $\rad$, a desired probability $1-\prttails$ and a chosen $\mathcal{K} \subset \mathbb{N}$, %
we optimize 
\begin{align*}
    \min_{\rtdesign} \max_{\rtvar \in \mathcal{K}}  \vert \entropy{\pdfest[\rtdesign]} - \entropy{\pdfest[\rtvar]} \vert.
\end{align*}
We make the following assumption on the first return time of an RW on an RRG, which can be separated into retroceding ($\retro$) and non-retroceding trajectories ($\nretro$). The distributions are known from \cite{Tishby_2021}.
\begin{assumption} \label{ass:frt}
The first return time on RRGs is dominated by non-retroceding scenarios, i.e., $\Pr(\tfr = t) = \Pr(\tfr = t \vert \nretro)$, which is described by $Pr(\tfr > t \vert \nretro)=  e^{-\cfac \frac{t-2}{\nnodes-2}}$ for $t\geq 3$ and $\Pr(\tfr > t \vert \nretro)=1$ otherwise.
\end{assumption}
We further have
$
    \expnretro = 2 + ({1-e^{-\cfac \frac{1}{\nnodes-2}}})^{-1}. 
$
By the choice of $\tdistrrg$ supported on $[\distmin, \distmax]$, each node selects the destination node in a distance of at least $\distmin$. Hence, at any destination node, the minimal observed return time is $\tfr \geq 2\distmin$ and hence the probability that a destination node observes a return time of $\tfr \geq t\geq 2\distmin$ is given by
\begin{align}
    \Pr(\tfr \geq t \vert \nretro, \tfr \geq 2\distmin)
    =e^{-\cfac \frac{t-2\distmin}{\nnodes-2}}. \label{eq:critical_cdf}
\end{align}

We justify \cref{ass:frt} by the fact that we are interested in first return times of at least $2\distmin$. In this case, the probability $\Pr(\nretro) = 1/(c-1)$ diminishes further, making $\Pr(\retro \vert \tfr \geq 2\distmin)$ the dominating component. With this at hand, we have the following main result of our paper that quantifies $\alpha$-privacy under a probabilistic guarantee.

\begin{theorem} \label{thm:optimalsol_entropybound}
    Let $\prttails>0$ and $d\define\distmax-\distmin$. Then, there exists a value for $\rtdesign$ such that with probability at least $1-\prttails$ for a certain $\mathcal{K}$ s.t. $\Pr(\rtvar \in \mathcal{K}) \geq 1-\prttails$, the entropy observed by every destination node $j$ is bounded by
    \begin{align*}
        \max_{\rtdesign} \min_{\rtvar \in \mathcal{K}} \entropy{\pdfest[\rtvar]} 
        \geq \underbrace{(1-\tveps) \log(d+1) + \tveps \log(\tveps) - \tveps}_{\alpha}, \\[-.8cm]
    \end{align*}
    where, for $\psp \define \Pr(\nsp\vert \distmax, \tfh \leq \rtvar)$,
    \begin{align*}
        \tveps = %
        O\left(\frac{1}{d} \frac{e^{\frac{\cfac}{2\nnodes} (\distmax+1) \psp \varepsilon} - e^{\frac{\cfac}{2\nnodes} \distmin \psp \varepsilon}}{e^{\frac{\cfac}{2\nnodes} \psp \varepsilon}-1} - 1\right)\!,
    \end{align*}
    and the asymptotic behavior of $\varepsilon$ is given by \\
    $\varepsilon \define \frac{\distmax}{2} O \left( \max\left\{(1-\prttails) e^{-(2\distmin-\distmax) \frac{\cfac}{\nnodes}} -1, \frac{-2-\log(1-\prttails)}{2\log(1-\prttails)} \right\} \right)$.
\end{theorem}

The result above is based on selecting an optimal value for $\rtdesign$ that minimizes the upper bound on the entropy. When optimal uniformity of $\pdfest[\rtvar]$ for $\rtvar=\rtdesign$ on $[\distmin, \distmax]$ should be achieved for the average return time, and hence $\rtdesign = \expnretro$, we have the following result. The result also holds for $\rtdesign \rightarrow \infty$, which corresponds to the case studied in \cref{sec:preliminary_solution_rrg}.

\begin{figure}[t]
    \centering
    \vspace{.1cm}
    \resizebox{.99\linewidth}{!}{\input{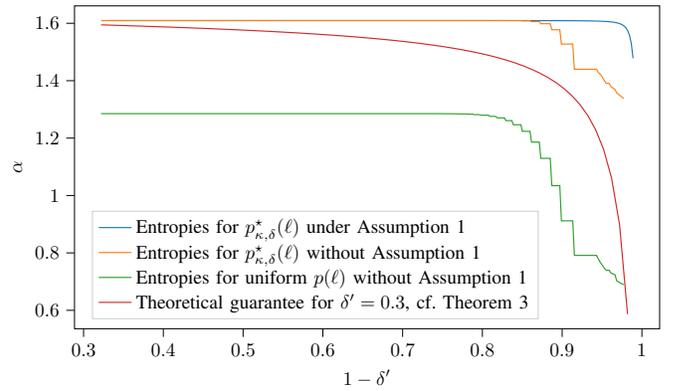}}
    \vspace{-0.25cm}
    \caption{$\alpha$-privacy over probabilistic guarantee for $\nnodes=300$, $c=3$ and $\rad=5$ with and without \cref{ass:fht}, compared with no countermeasure and the theoretical guarantee from \cref{thm:optimalsol_entropybound}.}
    \vspace{-.3cm}
    \label{fig:minimum_leakage_over_delta}
\end{figure}

\begin{corollary}
    When choosing $\tdistrrgopt$ to match the average case, i.e., for $\rtdesign = \expnretro$, and for $\rtdesign \rightarrow \infty$, \cref{thm:optimalsol_entropybound} holds with $\varepsilon$ replaced by $2\varepsilon$.
\end{corollary}

\section{Numerical Evaluations} \label{sec:experiments}
We evaluate our methods and theoretical guarantees on an RRG with $\nnodes=300$ nodes and degree $c=4$. Therefore, we consider the parameters $\rad=5$, $\prttails=0.3$, along with the choice of $\rtdesign=634$ and $\tdistrrgopt$ resulting from \cref{thm:optimalsol_entropybound} as well as $\distmin=2$ and $\distmax=6$. We approximate $\Pr(\nsp\vert \distmax, \tfh \leq \rtvar) \approx \Pr(\nsp\vert \distmax)$ for all $\rtvar$. In \cref{fig:leakage_over_sideinformation}, we compare the entropy in \cref{PrivacyNotion} for various side information $\rtvar$ under \cref{ass:fht} to the case where the optimal solution is found based on \cref{lemma:rrgopt} using the exact properties of RRGs, i.e., without the relaxation in \cref{ass:fht}. As a baseline, we plot the source node anonymity without any countermeasure, i.e., for choosing $\tdistrrgrad$ uniformly on $[\distmin, \distmax]$, and we also show our worst-case bound provided by  \cref{thm:optimalsol_entropybound}. 
For the same choice of $\rtdesign$, we plot in \cref{fig:minimum_leakage_over_delta} the minimum entropy resulting from \cref{thm:optimalsol_entropybound} as a function of the probability $1-\prttails$, illustrating how anonymity degrades as this probability varies. With approximately $ 90\%$ probability, our method still achieves near-optimal anonymity. 
Lastly, we analyze source node anonymity for $\rtvar=\rtdesign$ over the mean iteration time $\avgitertime = \sum_{\dist=\distmin}^{\distmax} \tdistrrgopt \mathbb{E}[\tfh\vert\dist]$ determined by the choice of $\tdistrrgopt$ and the average first hitting times $\mathbb{E}[\tfh \vert \dist]$. We observe an almost linear increase of $\alpha$ in $\avgitertime$ (cf. \cref{fig:mean_iteration_times} in \cref{app:privacy-utility-tradeoff}).

\section{Conclusion}
We considered the problem of privacy in decentralized random walk-based learning algorithms. Instead of resorting to only applying differential privacy guarantees, we formulated a new privacy notion based on revealing the model update, %
but concealing the identity of the owner of the revealed update. To that end, public key cryptography is used by the sender to encrypt the update with the public key of a designated destination, ensuring that no intermediate node can decrypt the model update. The choice of the destination is the key component. We designed a probability distribution over the choice of the destination that ensures that with high probability, the destination will not be able to guess the identity of the source. %

\bibliographystyle{IEEEtran}

\bibliography{references_paper}

\appendix
\subsection{Proof of \cref{lem:unnormalized_likelihood_sp} and \cref{lem:unnormalized_likelihood}}
\begin{proof}
We only prove \cref{lem:unnormalized_likelihood} as it includes \cref{lem:unnormalized_likelihood_sp} as a special case for $\rtvar\to\infty$. First note that with \cref{ass:fht}, for every $\rtvar>l$,
\begin{align*}
      &\int_{\mathbb{E}[\tfh(i,j) \vert \tfh \leq \rtvar]-\rad}^{\mathbb{E}[\tfh(i,j) \vert \tfh \leq \rtvar]+\rad} p_{D_i}(j) p_{i \rightarrow j} (t) dt\\
      &= \frac{\tdistrrg }{A(\dist)}P(\nsp\vert \dist)\hspace{-0.5cm}\sum_{t=\truncexp{\rtvar}-\rad}^{\truncexp{\rtvar}+\rad}\hspace{-0.4cm} P(\tfh=t\vert \dist,\nsp). 
\end{align*}
Inserting (\ref{Gl4}) yields 
\begin{align*}
    &\sum_{t=\truncexp{\rtvar}-\rad}^{\truncexp{\rtvar}+\rad} P(\tfh=t\vert \dist,\nsp)\\
    &= \left(e^{\frac{\cfac}{\nnodes}}-1\right) \sum_{t=\truncexp{\rtvar}-\rad}^{\truncexp{\rtvar}+\rad} e^{-\frac{\cfac (t-l)}{\nnodes}}\\
    &= \left(e^{\frac{\cfac}{\nnodes}}-1\right) \sum_{t=0}^{2\rad} e^{-\frac{\cfac (\truncexp{\rtvar}+t-\rad-l)}{\nnodes}}\\
    &= \left(e^{\frac{\cfac}{\nnodes}}-1\right)e^{\left(-\frac{\cfac (\truncexp{\rtvar}-\rad-l)}{\nnodes}\right)} \frac{e^{-\cfac\frac{2\rad+1}{N}}-1}{e^{-\frac{\cfac}{N}}-1}, 
\end{align*}
which finishes the proof.

\end{proof}

\subsection{Proof of \cref{lemma:rrgoptprelim_sp} and \cref{lemma:rrgopt}}

\begin{proof}
Again, we only prove \cref{lemma:rrgopt} as a generalization of \cref{lemma:rrgoptprelim_sp}. 
Assume that for some $\commonsol>0$, 
\begin{align}\label{Gl8}
    \tdistrrgopt \fullfacrt[\rtdesign] = \commonsol
\end{align}
is constant. Then with  $p_{\rtdesign, \rad}^\star$ being a probability distribution over $[\distmin,\distmax]$ defined as 
$ \tdistrrgopt =\frac{\commonsol}{\fullfacrt[\rtdesign] }$ we have 
\begin{align*}
    1=\sum_{\dist=\distmin}^{\distmax}\tdistrrgopt=\sum_{\dist=\distmin}^{\distmax} \frac{\commonsol}{\fullfacrt[\rtdesign]}
\end{align*}
or equally, 
\begin{align*}
    s = \frac{1}{\sum_{\dist=\distmin}^{\distmax} \frac{1}{\fullfacrt[\rtdesign]}}. 
\end{align*}
This proves the representation of $\tdistrrgopt$ in (\ref{Gl8}). The uniform distribution of $\pdfest[\rtdesign]$ over the support $[\distmin,\distmax]$ follows trivially.

\end{proof}

\subsection{Proof of \cref{prop:truncexpfht}}

\begin{proof} %
To compute the conditional expectation $\truncexpnsp{\rtvar} $, note that
\begin{align*}
    \truncexpnsp{\rtvar} 
    &=\frac{\sum_{t = 0}^{\rtvar} P(\tfh>t|\dist, \nsp)}{\Pr(\tfh \leq \rtvar|\dist, \nsp)}.
\end{align*}
According to (\ref{Gl4}), the tail of the first hitting time distribution is given by
\begin{align*}
    \Pr(\tfh > t|\dist, \nsp) = \begin{cases}
        1, & t\leq \dist \\
     e^{-\frac{\cfac}{\nnodes} (t-\dist)}, & t > \dist.
    \end{cases} 
\end{align*}
Hence, we have
\begin{align*}
    &\sum_{t = 0}^{\rtvar} P(\tfh>t|\dist, \nsp) = \dist + \sum_{t=\dist}^{\rtvar} e^{-\frac{\cfac}{\nnodes}(t-\dist) } \\
    &= \dist + \sum_{t^\prime=0}^{\rtvar-\dist} e^{-\frac{ \cfac}{\nnodes}t^\prime} = \dist + \frac{e^{-\frac{\cfac}{\nnodes} (\rtvar-\dist)}-1}{e^{-\frac{\cfac}{\nnodes}}-1}
\end{align*}
and 
\begin{align*}
    \truncexpnsp{\rtvar}
    &= \frac{\dist}{1 - e^{-\frac{\cfac}{\nnodes} (\rtvar-\dist)}} + \frac{1}{1-e^{-\frac{\cfac}{\nnodes}}},
\end{align*}
whereby the second term is independent of $\kappa$. Note that for $\kappa \rightarrow \infty$, we obtain the known case in (\ref{Gl3}) as  
\begin{align*}
    &\expnsp = \frac{\dist + \frac{e^{-\frac{\cfac}{\nnodes} (\rtvar-\dist)}-1}{e^{-\frac{\cfac}{\nnodes}}-1}}{1 - e^{-\frac{\cfac}{\nnodes} (\rtvar-\dist)}} = \dist + \frac{1}{1-e^{-\frac{\cfac}{\nnodes}}}. 
\end{align*}

\end{proof}

\subsection{Proof of \cref{thm:tvboundest}}

\begin{proof}
We first derive the following intermediate result on the unnormalized likelihoods:
\begin{align*}
&\vert \tdistrrg \fullfacrt[\rtvar] - \tdistrrg \fullfacrt[\rtdesign] \vert \\
&= \partfac \tdistrrg \left \vert e^{-\frac{\cfac \truncexp{\rtvar}}{\nnodes}} - e^{-\frac{\cfac \truncexp{\rtdesign}}{\nnodes}} \right \vert \\
&\leq \begin{aligned}[t] \partfac\tdistrrg e^{-\frac{\cfac \truncexp{\rtdesign}}{\nnodes}} \errboundtmp\end{aligned}, 
\end{align*}
where 
\begin{align*}
        \errboundtmp &\define e^{\frac{\cfac \vert \truncexp{\rtvar} - \truncexp{\rtdesign} \vert}{\nnodes}} - 1.
\end{align*}
using that $\vert e^{-x} - 1 \vert \leq e^{\vert x \vert} - 1,$ for all $x \in \mathbb{R}$. According to Lemma \ref{lem:unnormalized_likelihood}, this equals 
\begin{align}\label{Gl9}
    &\vert \tdistrrg \fullfacrt[\rtvar] - \tdistrrg \fullfacrt[\rtdesign] \vert \notag\\
    &\leq \fullfacrt[\rtdesign] \tdistrrg \errboundtmp = \commonsol \errboundtmp,
\end{align}
with $s$ from (\ref{Gl8}). 
For every $\rtvar, \rtdesign>0$, let 
\begin{align*}
    &\errbound \define \\
    &\left \vert \truncexpnsp{\rtvar} - \truncexpnsp{\rtdesign} \right \vert. 
\end{align*}
According to \cref{prop:truncexpfht}, $\errbound$ is given by 
\begin{align*}
    \errbound = \dist \left|\frac{1}{1 - e^{\cfac/\nnodes (\rtvar-\dist)}}-\frac{1}{1 - e^{-\cfac/\nnodes (\rtdesign-\dist)}}\right|
\end{align*}
and with \cref{ass:eth}, we can express $\errboundtmp$ as
\begin{align*}
     \errboundtmp = e^{\frac{\cfac \Pr(\nsp \vert \dist, \tfh \leq \rtvar) \errbound}{N}}-1.
\end{align*}
Let $\diam=\distmax-\distmin$ be the number of possible distances and recall from (\ref{Gl8}), that $  = \tdistrrg \fullfacrt[\rtdesign]\commonsol$ is constant for some $\commonsol>0$. To bound the total variation between $\pdfest$ and $\pdfest[\rtdesign]$, from the definition of total variation, we have
\begin{align*}
    &\tv{\pdfest}{\pdfest[\rtdesign]} \\
    &= \frac{1}{2} \sum_{\dist} \left\vert \frac{\tdistrrg \fullfacrt[\rtvar]}{\sum_{\dist^{\prime}} \tdistrrg[\dist^{\prime}] \fullfacrtprime[\rtvar]} - \frac{\tdistrrg \fullfacrt[\rtdesign]}{\sum_{\dist^{\prime}} \tdistrrg[\dist^{\prime}] \fullfacrtprime[\rtdesign]} \right\vert \\
    &\leq   \frac{1}{2} \sum_{\dist}  A_{\delta,\kappa,\kappa^{\prime}}(\dist) + B_{\delta,\kappa,\kappa^{\prime}}(\dist), 
\end{align*}
where 
\begin{align*}
    A_{\delta,\kappa,\kappa^{\prime}}(\dist)\define \left\vert \frac{\tdistrrg \fullfacrt[\rtvar]}{\sum_{\dist^{\prime}} \tdistrrg[\dist^{\prime}] \fullfacrtprime[\rtvar]} - \frac{\tdistrrg  \fullfacrt[\rtvar]}{\diam \commonsol}\right\vert
\end{align*}
and 
\begin{align*}
    B_{\delta,\kappa,\kappa^{\prime}}(\dist)\define \left\vert\frac{\tdistrrg  \fullfacrt[\rtvar]}{\diam \commonsol} - \frac{\tdistrrg  \fullfacrt[\rtdesign]}{\diam \commonsol} \right\vert,
\end{align*}
according to the triangle inequality. Note that using the triangle inequality once again and in view of (\ref{Gl9}), we have
\begin{align*}
    A_{\delta,\kappa,\kappa^{\prime}}(\dist) &\leq  \frac{\tdistrrg \fullfacrt[\rtvar]}{\diam\commonsol}  \frac{\left\vert \sum_{\dist^{\prime}} \tdistrrg[\dist^{\prime}] \fullfacrtprime[\rtvar] - \diam \commonsol \right\vert}{\sum_{\dist^{\prime}} \tdistrrg[\dist^{\prime}] \fullfacrtprime[\rtvar]}\\
    & \leq  \frac{\tdistrrg \fullfacrt[\rtvar]}{\diam\commonsol}  \frac{\sum_{\dist^{\prime}} \vert\tdistrrg[\dist^{\prime}] \fullfacrtprime[\rtvar] - \tdistrrg[\dist^{\prime}] \fullfacrtprime[\rtdesign]\vert}{\sum_{\dist^{\prime}} \tdistrrg[\dist^{\prime}] \fullfacrtprime[\rtvar]}\\
    &\leq \frac{\tdistrrg \fullfacrt[\rtvar]}{\diam} \frac{\sum_{\dist^{\prime}} \errboundtmpprime}{\sum_{\dist^{\prime}} \tdistrrg[\dist^{\prime}] \fullfacrtprime[\rtvar]}
\end{align*}
and hence, 
\begin{align*}
  \sum_{\dist}  A_{\delta,\kappa,\kappa^{\prime}}(\dist) &\leq  \frac{\sum_{\dist^{\prime}} \errboundtmpprime}{\diam} \sum_{\dist}\frac{\tdistrrg \fullfacrt[\rtvar]}{\sum_{\dist^{\prime}} \tdistrrg[\dist^{\prime}] \fullfacrtprime[\rtvar]}\\
  &= \frac{\sum_{\dist^{\prime}} \errboundtmpprime}{\diam}. 
\end{align*}
Similarly, we obtain 
\begin{align*}
     B_{\delta,\kappa,\kappa^{\prime}}(\dist) = \frac{\errboundtmp}{d}
\end{align*}
such that 
\begin{align*}
      &\tv{\pdfest}{\pdfest[\rtdesign]} \\
      &\leq \frac{1}{2} \left(\frac{\sum_{\dist} \errboundtmp}{\diam} + (\frac{\sum_{\dist} \errboundtmp}{\diam}\right) 
      = \frac{1}{\diam}\sum_{\dist} \errboundtmp.
\end{align*}
Observing that
\begin{align*}
\sum_\dist \errboundtmp  &= \sum_\dist \left( e^{\frac{\cfac \Pr(\nsp \vert \dist, \tfh \leq \rtvar) \errbound}{N}}-1\right)
\end{align*}
concludes the proof.

\end{proof}

\subsection{Proof of \cref{thm:estentropybound}}

\begin{proof}
From \cite{zhang2007estimating} and an alphabet size of $\distmax - \distmin+1$, we have
\begin{align*}
    \vert \entropy{\pdfest[\rtdesign]} - \entropy{\pdfest[\rtvar]} \vert \leq \tveps \log(\distmax-\distmin) + \binentropy{\tveps},
\end{align*}
where $\binentropy{\tveps} \define - \tveps \log(\tveps) - (1-\tveps) \log(1-\tveps)$ is the binary entropy function. 
From \cref{lemma:rrgopt}, it is known that $\pdfest[\rtdesign]$ is uniform over $[\distmax-\distmin]$, and hence $\entropy{\pdfest[\rtdesign]} = \log(\distmax-\distmin+1)$. For the binary entropy function, it holds that
\begin{align*}
    \binentropy{\tveps} \leq - \tveps \log(\tveps) + \tveps,
\end{align*}
and hence, by slightly relaxing the above inequality, that
\begin{align*}
    \vert \entropy{\pdfest[\rtdesign]} - \entropy{\pdfest[\rtvar]} \vert \leq \tveps \log(\distmax-\distmin+1) - \tveps \log(\tveps) + \tveps.
\end{align*}
The statement follows since $\entropy{\pdfest[\rtdesign]} = \log(\distmax-\distmin+1)$.
\end{proof}

\subsection{Proof of \cref{thm:optimalsol_entropybound}}

\begin{proof}
From \cref{ass:frt} and (\ref{eq:critical_cdf}), the probability that $\tfr$ is between $t_1 > 1$ and $t_2 > t_1$ is given by
\begin{align*}
    \Pr(&t_1 \leq \tfr \leq t_2 \vert \tfr \geq 2\distmin) \\
    &=e^{-\cfac \frac{t_1-2\distmin}{\nnodes-2}} - e^{-\cfac \frac{t_2+1-2\distmin}{\nnodes-2}} \\
    &\geq e^{-\cfac \frac{t_1-2\distmin}{\nnodes}} - e^{-\cfac \frac{t_2+1-2\distmin}{\nnodes}} \\
    &= e^{2\cfac \frac{\distmin}{\nnodes}} \left(e^{-\cfac \frac{t_1}{\nnodes}} - e^{-\cfac \frac{\rtdesign}{\nnodes}} + e^{-\cfac \frac{\rtdesign}{\nnodes}} - e^{-\cfac \frac{t_2+1}{\nnodes}}\right),
\end{align*}
where the inequality follows from the monotonicity of $e^{-x}$ and as $t_2 > t_1$. To achieve a probability of at least 
\begin{align*}
    \Pr(t_1 \leq \tfr \leq t_2 \vert \tfr \geq 2\distmin) \geq 1-\prttails, 
\end{align*}
since deviations are maximal for small $t_1$, we choose $t_2 = \infty$. Hence, we require that $t_1 \leq -\frac{\nnodes}{\cfac} \log(1-\prttails) + 2\distmin$.
Motivated by \cref{thm:tvboundest}, we choose $t_1$ and $t_2$ so that
\begin{align*}
    e^{-\cfac \frac{t_1}{\nnodes}} - e^{-\cfac \frac{\rtdesign}{\nnodes}} = e^{-\cfac \frac{\rtdesign}{\nnodes}} - e^{-\cfac \frac{t_2}{\nnodes}} = \frac{1-\prttails}{2e^{2\distmin \cfac/\nnodes}}.
\end{align*}
Consider \cref{thm:tvboundest}, and let $t_1 \leq \rtvar \leq t_2$. Then, by continuity for both cases of $\rtdesign > \rtvar$ and $\rtdesign < \rtvar$, the worst-case bound is given for $\rtvar=t_1$ and $\rtvar=t_2$. By the deliberate choice of $t_1$ and $t_2$ above, choosing $\rtvar \in \{t_1, t_2\}$ gives the same result. The resulting probability that $\tfr$ is between $t_1$ and $t_2$ is at least $1-\prttails$. With $t_2 = \infty$, we choose $t_1 = -\frac{\nnodes}{\cfac} \log(1-\prttails) + 2\distmin$. We ignore rounding errors since we are interested in order-wise results. There exists a choice of $\rtdesign$ such that $\errbound[t_2] = \errbound[t_1] = \frac{\dist}{2} \left\vert \frac{1}{1 - e^{-\cfac/\nnodes (t_1-\dist)}}-\frac{1}{1 - e^{-\cfac/\nnodes (t_2-\dist)}}\right\vert = \frac{\dist}{2} \left(\frac{1}{1 - e^{-\cfac/\nnodes (t_1-\dist)}}-1\right)$, for
\begin{align*}
    \errbound[t_2] &= \dist \left\vert \frac{1}{1 - e^{-\cfac/\nnodes (t_2-\dist)}}-\frac{1}{1 - e^{-\cfac/\nnodes (\rtdesign-\dist)}}\right\vert
\end{align*}
and 
\begin{align*}
    \errbound[t_1] &= \dist \left\vert \frac{1}{1 - e^{-\cfac/\nnodes (t_1-\dist)}}-\frac{1}{1 - e^{-\cfac/\nnodes (\rtdesign-\dist)}}\right\vert.
\end{align*}
For small $1-\prttails$, we approximate $\frac{1}{1-e^{x}}$ by $\frac{1}{x} + \frac{1}{2}$, and obtain
\begin{align*}
    \errbound[t_1] &= \errbound[t_2] = \frac{\dist}{2} \left( \frac{1}{1 - e^{-\cfac/\nnodes (t_1-\dist)}} - 1\right) \\
    &= O\left( \frac{\dist}{2} \left(\frac{\nnodes}{\cfac (t_1-\dist)} - \frac{1}{2} \right) \right) \\
    &= O\left( \frac{\dist}{2} \left(\frac{1}{-\log\left(1-\prttails\right) + (2\distmin-\dist) \frac{\cfac}{\nnodes}} - \frac{1}{2}\right) \right) \\
    &= \dist O\left(\frac{1}{-2\log\left(1-\prttails\right)} - \frac{1}{4} \right).
\end{align*}
For large $1-\prttails$, we approximate $\frac{1}{1-e^{x}}$ by $e^{-x}$ and obtain
\begin{align*}
    \errbound[t_1] &= \errbound[t_2] = O \left(\frac{\dist}{2} e^{-\cfac/\nnodes (t_1-\dist)} - 1 \right)  \\
    &= O\left( \frac{\dist}{2} e^{\log\left(1-\prttails\right) - (2\distmin-\dist) \frac{\cfac}{\nnodes}} - 1 \right) \\
    &= O\left( \frac{\dist}{2} (1-\prttails) e^{-(2\distmin-\distmax) \frac{\cfac}{\nnodes}} -1 \right) 
\end{align*}
Hence, overall we have for $t_1 \leq \rtvar \leq t_2$ that
$\errbound[\rtvar]=\frac{\dist}{2}\epsilon$ with \\
$$\epsilon \define O \left( \max\left\{(1-\prttails) e^{-(2\distmin-\distmax) \frac{\cfac}{\nnodes}} -1, \frac{-1}{\log(1-\prttails)}-\frac{1}{2} \right\} \right)\!.$$ 
This formulation of $\errbound[\rtvar]$ is now independent of $\dist$. According to \cref{thm:tvboundest}, the total variation is bounded as
\begin{align*}
    \tv{\pdfest}{\pdfest[\rtdesign]} &\leq \frac{1}{\distmax-\distmin} \sum_{\dist=\distmin}^{\distmax} e^{\frac{\cfac}{\nnodes} \psp\errbound} - 1 \\
    &= \frac{1}{\distmax-\distmin} \sum_{\dist=\distmin}^{\distmax} e^{\frac{\cfac\psp}{\nnodes} \frac{\dist}{2} \varepsilon} - 1 \\
    &= O\left(\frac{1}{\distmax-\distmin} \frac{e^{\frac{\cfac\psp}{2\nnodes} (\distmax+1) \varepsilon} - e^{\frac{\cfac\psp}{2\nnodes} \distmin \varepsilon}}{e^{\frac{\cfac\psp}{2\nnodes} \varepsilon}-1} - 1\right)\!.%
\end{align*}
This concludes the proof.

\end{proof}

\subsection{Privacy-Utility Trade-Off} \label{app:privacy-utility-tradeoff}

For completeness, we provide in \cref{fig:mean_iteration_times} a visual representation of the trade-off between the privacy parameter $\alpha$ and the mean per-iteration time $\avgitertime$ for $\nnodes=300$ and $c=4$.
\begin{figure}[ht]
    \centering
    \resizebox{\linewidth}{!}{\begin{tikzpicture}

\definecolor{color0}{rgb}{0.12156862745098,0.466666666666667,0.705882352941177}

\begin{axis}[
legend style={fill opacity=0.8, draw opacity=1, text opacity=1, draw=white!80!black},
tick align=outside,
tick pos=left,
x grid style={white!69.0196078431373!black},
xlabel={\(\displaystyle T_{\ell_1, \ell_2}\)},
xmin=533.958100283192, xmax=587.609587358798,
xtick style={color=black},
y grid style={white!69.0196078431373!black},
ylabel={\(\displaystyle \mathrm{H}\left(W_{\delta, \kappa}\right)\)},
ymin=1.04957082601752, ymax=2.12848300433042,
ytick style={color=black}
]
\addplot [semithick, color0, forget plot]
table {%
536.396804241174 1.09861228866811
559.023513899782 1.38629436111989
572.833580896962 1.6094379124341
580.571769233086 1.79175946922805
584.168325899114 1.94591014905531
585.170883400816 2.07944154167984
};
\end{axis}

\end{tikzpicture}}
    \caption{Entropy over the average iteration times $\avgitertime$ for $\rtvar=\rtdesign$ as chosen in \cref{sec:experiments}, $\nnodes=300$ and $c=4$.}
    \label{fig:mean_iteration_times}
\end{figure}

\end{document}